\documentclass[final] {aipproc}

\layoutstyle{6x9}

\begin{document}

\title{Gamma-ray Bursts in Wavelet Space}

\author{Zsolt Bagoly}{
address={Laboratory for Information Technology, E\"{o}tv\"{o}s University,
H-1117 Budapest, P\'azm\'any P. s.  1./A, Hungary}}
\author{Istv\'an Horv\'ath}{
address={Department of Physics, Bolyai Military University, H-1456 Budapest,
POB 12, Hungary}}
\author{Attila M\'esz\'aros}{
address={Astronomical Institute of the Charles University, V
Hole\v{s}ovi\v{c}k\'ach 2, CZ-180 00 Prague 8, Czech Republic},
altaddress={Stockholm Observatory, AlbaNova, SE-106 91 Stockholm, Sweden} }
\author{Lajos G. Bal\'azs}{
address={Konkoly Observatory, H-1525 Budapest, POB 67, Hungary}}

\begin{abstract}
The gamma-ray burst's lightcurves have been analyzed using a special wavelet
transformation.  The applied wavelet base is based on a typical Fast
Rise-Exponential Decay (FRED) pulse.  The shape of the wavelet coefficients'
total distribution is determined on the observational frequency grid.  Our
analysis indicates that the pulses in the long bursts' high energy channel
lightcurves are more FRED-like than the lower ones, independently from the
actual physical time-scale.  
\end{abstract}

\maketitle

\section{Introduction}

The shape of the gamma-ray burst's (GRB's) $64$ ms resolution lightcurves in
the BATSE Gamma-Ray Burst Catalog \cite{Meeganetal2000} carry an immense amount
of information.  However, the chaging S/N ratio complicates the detailed
comparative analysis of the lightcurves.  During the morphological analysis of
the GRB's \cite{ 1992como.work...61K, 1994AAS...185.1507N, 1998hgrb.symp..171N}
a subclass with Fast Rise-Exponential Decay (FRED) pulse shape were
observed.  This shape is quite attractive because its fenomenological
simplicity.  Here we use a special wavelet transformation with a kernel
function based on  a FRED-like pulse.  Similar approach have been used by
\cite{2002AA...385..377Q}, but their base functions were constructed
differently.

\section{The FRED Wavelet Transform}

We have used the Discrete Wavelet Transform (DWT)
matrix formalism (e.g. \cite{press}): here for an input data vector $v$, the
one step of the wavelet transform is a multiplication with a special matrix
$F$:
\begin{equation}
\mathbf{F}=\left[\matrix{
 c_0 &  c_1 &  c_2 &  c_3 &&&&&&&\cr
 c_3 & -c_2 &  c_1 & -c_0 &&&&&&&\cr
&& c_0 &  c_1 &  c_2 &  c_3 &&&&&\cr
&& c_3 & -c_2 &  c_1 & -c_0 &&&&&\cr
\vdots & \vdots &&&&& \ddots &&&&\cr
 c_2 &  c_3 &&&&&&&&  c_0 &  c_1 \cr
 c_1 & -c_0 &&&&&&&&  c_3 & -c_2 \cr
}\right]
\end{equation}
where the $c_0,\ldots,c_3$ are the 4-stage FIR filter parameters defining the
wavelet.  To obtain these values we require the matrix $F$ to be orthogonal
(e.g. no information loss), and the output of the even (derivating-like) rows
should disappear for a constant and for a FRED-like $e^{-\mbox{t}/\tau}$ input
signal.  These requirements give two different solutions for $c_0,\ldots,c_3$:
a rapidly oscillating one and a smooth one. In the following we'll use the
later one.

Our filter process with the FRED wavelet transform consists of the usual
$\mbox{{\em transform}} \rightarrow \mbox{{\em filter/cut}} \rightarrow \mbox{{\em inverse transform}}$
digital filtering steps.  During the filtering we'll loose some information,
however this could be quite small. To demonstrate the efficiency of the
algorithm on Fig.~1. we reconstructed the 100-320 keV 64ms ligthcurve of BATSE
trigger 0143 from the biggest 5\% of the total wavelet coefficients. 
The excellent reconstruction of each individual pulse is obvious.

The wavelet transformation algorithm divides the phase-space into equal area
regions.  On Fig.~2. the wavelet transform are shown. Here the dark segments
are the really important coefficients - however they cover only a small portion
of the total area which explains the high efficiency of the reconstruction.

\begin{figure}
\includegraphics[width=0.6\textwidth]{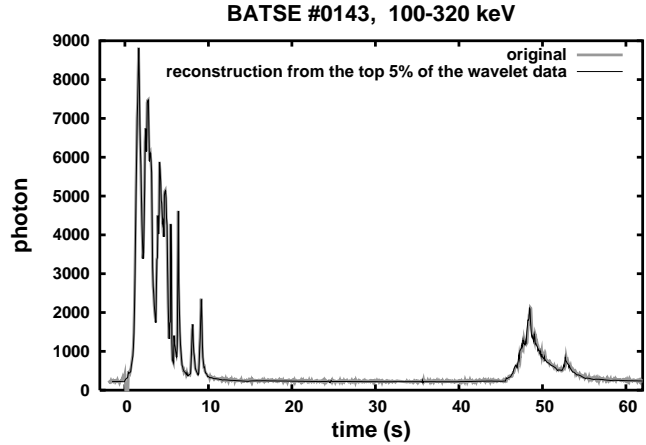}
\caption{The original and the reconstructed ligthcurve of BATSE trigger 0143
- only 5\% of the total wavelet data is  used. }
\end{figure}

\begin{figure}
\includegraphics[width=0.54\textwidth]{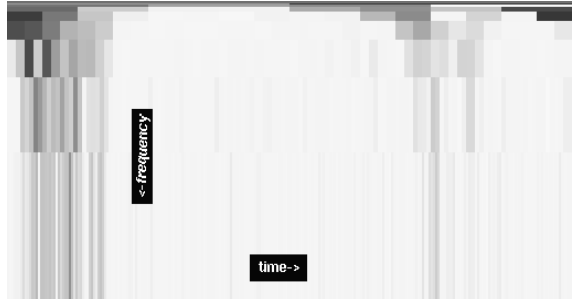}
\caption{The wavelet phase-space density for BATSE trigger 0143.}
\end{figure}

\section{Wavelet scale analysis}

%On Fig.~2. the squares denote the different time-frequency cells. 
For a frequency-like wavelet scale analysis we would like to create a
power-spectrum like distribution along the frequency axis. However, one should
be careful.  In the classical signal processing one uses the power spectrum
from the Fourier-transform, because  the signals are electromagnetic-like
usually, e.g.  the power (or energy) is proportional to the  square of the
signal. Here the lightcurves measure photon counts --- so the signal's energy
is simply the sum of the counts. For this reason we approximate the signal's
strength as a sum the magnitude of the coefficients along the given frequency
rows.

\begin{figure}
\includegraphics[width=0.45\textwidth]{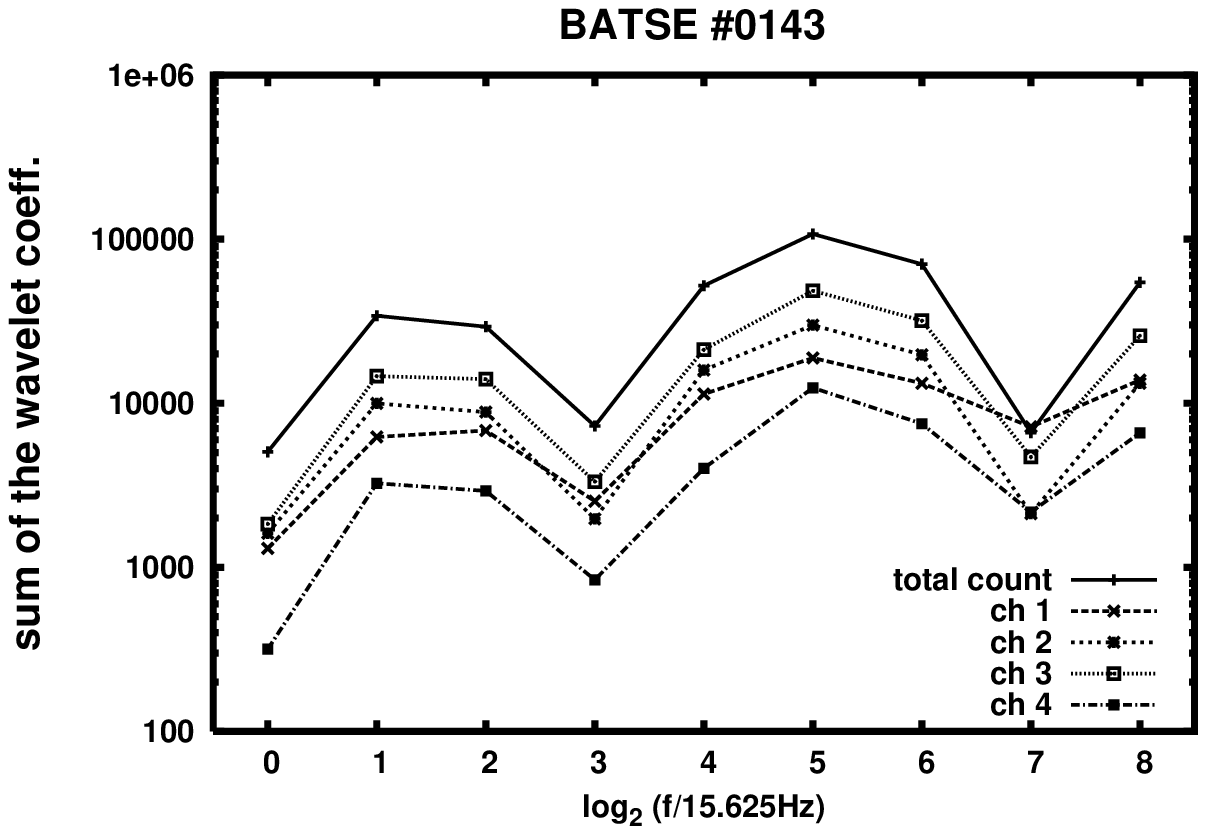}
\includegraphics[width=0.45\textwidth]{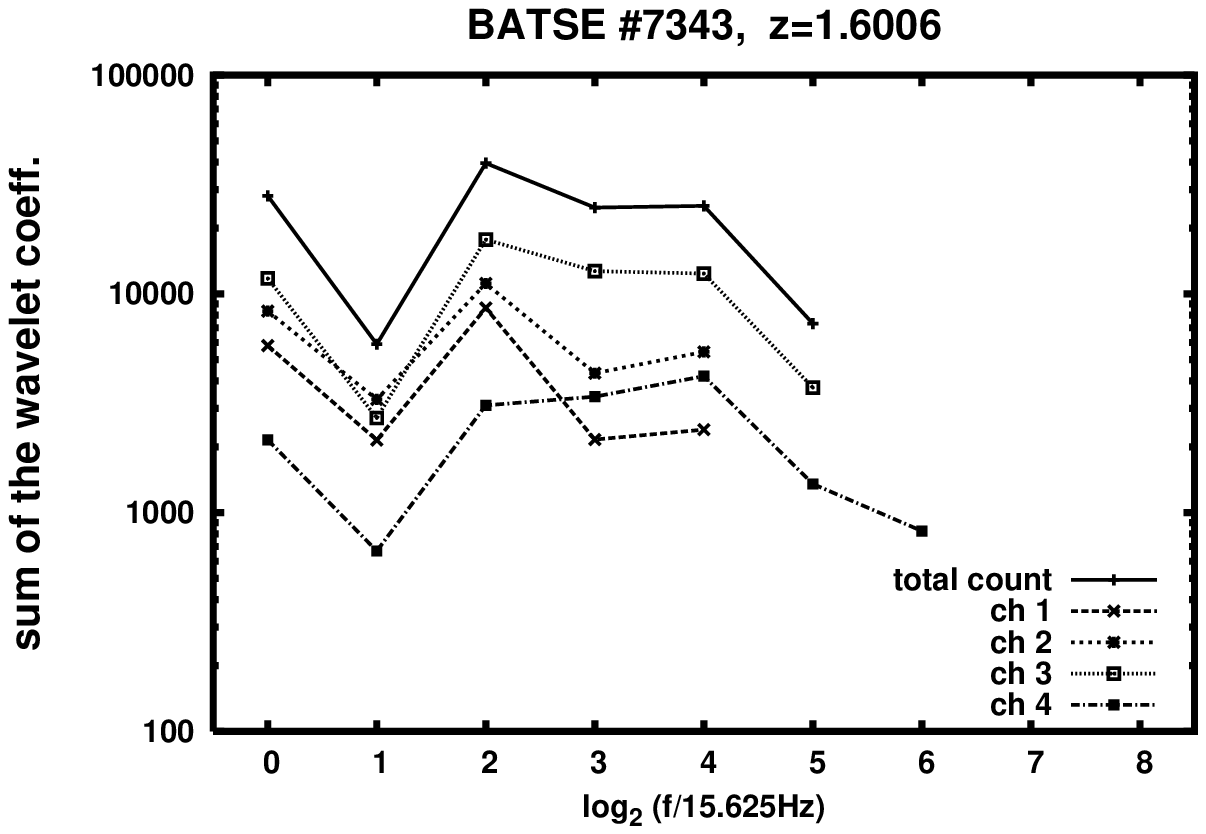}
\caption{The wavelet signal's frequency distribution for BATSE triggers 0143
and 7343 respectively.} 
\end{figure}

This signal's strength indicate on Fig~3. (BATSE trigger 0143) the 
maximum power to be around $f \approx 500
\mbox{Hz} $. In each energy channel the signals are similar (observe the
logarithmic scale), because the signal is strong even at high energies (channel
4).  For BATSE trigger 7343 (with optical redshift $z=1.6006$) one can observe a strong
high frequency cutoff: some of the signal's high frequency part is missing.
However all the 4 channels are visible, while the maximum power is around $f
\approx 62.5 \mbox{Hz}$.  It is interesting to remark  that the signal's shape is
quite similar to trigger 0143 if that is scaled down by a factor of $\approx
2.4-2.8$ in frequency.

\section{Wavelet filtering and similarity}

The FRED wavelet transform measures the similarity between the different
wavelet kernel functions (here all are FRED-based) and the actual signal. To
quantify the similarity we define a magnitude cutoff in the wavelet space so,
that the {\em reconstructed} $T_{50}$ value from the filtered data should be
similar to the original values.  The $T_{50}$ value and its $\sigma_P$ error
from the photon count statistics could be easily determined from the original
ligthcurve.  To keep only the important features we define the $ T_{50
\mbox{break}} $ breakpoint where 
$$
 \mid T_{50 \mbox{break}} - T_{50} \mid = 10.0 \sigma_P  
$$
Using a cut-off point it is possible to define a Compressed Size (CS)
for a burst: it is the number of bins (in the wavelet space) needed to restore
the curve at the break.

The CS value is a robust measure quantifying the similarity between the FRED
kernel and the different channels' lightcurves.  Our analysis suggest that all
the low energy channels \#1, \#2 and \#3 behaves similarly, while the high
energy ($ >320$ keV) channel is different (which is not very surprising, e.g.
\cite{bag98}).  Fig.~4. shows the {\em ratio} of the CS's against the total
count lightcurves' CS for channels 2 and 4.  This distributions indicate that
the pulse-shapes in the long bursts' high energy channel are more FRED-like
than the lower ones - and this is {\em independent} from the actual FRED
time-scale!

\begin{figure}
\includegraphics[height=8.5cm,angle=270]{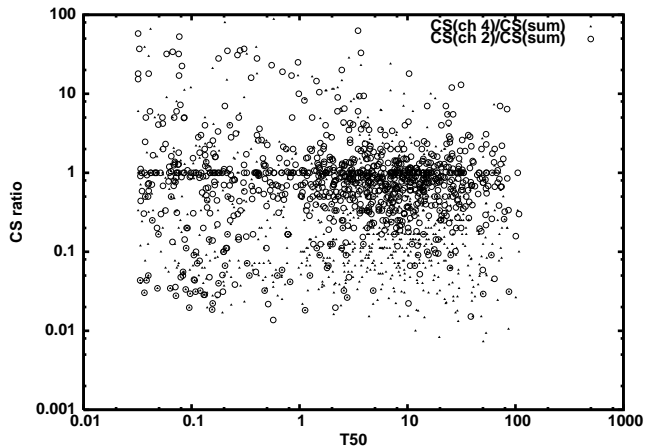}
\caption{The relative value of the CS's against the total lightcurves' CS 
for channels 2 and 4.}
\end{figure}

\begin{theacknowledgments}
This research was supported in part through OTKA grants T024027 (L.G.B.),
and T034549, Czech Research Grant J13/98: 113200004 
and by a grant from the Wenner-Gren Foundations (A.M.).
\end{theacknowledgments}

\bibliographystyle{aipproc}

\begin{thebibliography}{}

\bibitem[Bagoly et al. (1998)]{bag98} Bagoly, Z., M\'esz\'aros, A., Horv\'ath,
I., Bal\'azs, L. G., \& M\'esz\'aros, P. 1998, ApJ, 498, 342.

\bibitem[Kouveliotou et al. (1993)]{Kouveliotouetal1993} Kouveliotou, C.,
Meegan, C.A. \& Fishman, G.J. 1993, ApJ, 413, L101

\bibitem[Kouveliotou et al. (1992)]{1992como.work...61K} Kouveliotou, C.,
Paciesas, W.~S., Fishman, G.~J., Meegan, C.~A., \& Wilson, R.~B.\ 1992, The
Compton Observatory Science Workshop, 61
%editor

\bibitem[Meegan et al. (2000)]{Meeganetal2000} Meegan, C., Malozzi, R.S., Six,
F. \& Connaughton, V. 2001, Current BATSE Gamma-Ray Burst Catalog,
http://gammaray.msfc.nasa.gov/batse/grb/catalog

\bibitem[Norris et al. (1994)]{1994AAS...185.1507N} Norris, J.~P., Nemiroff,
R.~J., Bonnell, J.~T., Paciesas, W.~S., Kouveliotou, C., Fishman, G.~J., \&
Meegan, C.~A.\ 1994, American Astronomical Society Meeting, 26, 1333
%editor

\bibitem[Norris, Scargle, Bonnell, \& Nemiroff (1998)]{1998hgrb.symp..171N}
Norris, J.~P., Scargle, J.~D., Bonnell, J.~T., \& Nemiroff, R.~J.\ 1998,
Gamma-Ray Bursts, 4th Hunstville Symposium, 171
%editor

\bibitem[Press et al. (1992)]{press} Press W.H., Teukolsky S.A., Vetterling
W.T., Flannery B.P. 1992, Numerical Recipes in Fortran, Second Edition, Cambridge
University Press, Cambridge

\bibitem[Quilligan et al. (2002)]{2002AA...385..377Q} {{Quilligan}, F.,
{McBreen}, B., {Hanlon}, L., {McBreen}, S., {Hurley}, K.~J. \& {Watson}, D.},
2002, A\&A,  385, 377

\end{thebibliography}

\end{document}